\documentclass{article}
\usepackage[utf8]{inputenc}
\usepackage{url}
\usepackage[backend=biber]{biblatex}
\usepackage{xcolor}
\usepackage{graphicx}
\usepackage{enumerate}
\usepackage{array}
\usepackage{ragged2e}
\usepackage{amsmath}
\usepackage{appendix}
\usepackage{authblk}
\usepackage{mathtools}
\usepackage{mathptmx}
\usepackage{mathrsfs}
\usepackage{amssymb}
\usepackage{amsmath}
\usepackage{amssymb}
\usepackage{csquotes}
\usepackage{hyperref}
\title{The Opportunity to Regulate Cybersecurity in the EU (and the World): Recommendations for the Cybersecurity Resilience Act}

\author[1]{Kaspar Rosager Ludvigsen}
\author[2]{Shishir Nagaraja} 

\affil[1]{Department of Computer and Information Sciences, University of Strathclyde, kaspar.rosager-ludvigsen@strath.ac.uk}
\affil[2]{Department of Computer and Information Sciences, University of Strathclyde, shishir.nagaraja@strath.ac.uk}
\date{May 2022}

\begin{document}

\begin{abstract}

Safety is becoming cybersecurity under most circumstances. This should be reflected in the Cybersecurity Resilience Act whenever it is proposed and agreed upon in the European Union. In this paper, we define a range of principles which this future Act should build upon, a structure and argue why it should be as all encompassing as possible. We do this on the basis of what the cybersecurity research community for long have asked for, and on what constitutes clear hard legal rules instead of soft. Important areas such as cybersecurity should be taken seriously, by regulating it in the same way we see other types of critical infrastructure and physical structures, and be uncompromising and logical, to encompass the risks and potential for chaos which its ubiquitous nature entails. 

We find that principles which regulate cybersecurity systems' life-cycles in detail are needed, as is clearly stating what technology is being used, due to Kirkhoffs principle, and dismissing the idea of technosolutionism. Furthermore, carefully analysing risks is always necessary, but so is understanding when and how the systems manufacturers may fail or almost fail, all of these details must be expected and detailed. We do this through the following principles:

Ex ante and Ex post assessment, Safety and Security by Design, Denial of Obscurity, Dismissal of Infallibility, Systems Acknowledgement, Full Transparency, Movement towards a Zero-trust Security Model, Cybersecurity Resilience, Enforced Circular Risk Management, Dependability, Hazard Analysis and mitigation or limitation, liability, A Clear Reporting Regime, Enforcement of Certification and Standards, Mandated Verification of Security and Continuous Servicing.

To this, we suggest that the Act employs similar authorities and mechanisms as the GDPR, and create strong national authorities to coordinate inspection and enforcement in each Member State, with ENISA being the top and coordinating organ. 
\thanks{We thank Professor Angela Daly for her feedback.}    
\end{abstract}

\maketitle

\emph{EARLY DRAFT}

\section{Introduction}





Contrary to this, on 16 of March 2022, the European Commission launched a Call for Evidence for the a potential future `Cyber Resilience Act' (CRA).\footnote{See \url{https://ec.europa.eu/info/law/better-regulation/have-your-say/initiatives/13410-Cyber-resilience-act-new-cybersecurity-rules-for-digital-products-and-ancillary-services_en}, last accessed 5 May 2022.} The specific call concerns itself with creating an Impact Assessment, which is standard procedure before having any (or no) legislative changes in the European Union, but the ideas pitched at this early stage by the Commission are very intriguing and clearly positive.\footnote{Documentation for this can be found at \url{https://ec.europa.eu/6fc8c99b-20bf-48b3-9c39-5d049adcb1c6}, last accessed 5 May 2022.} 

This paper sets out to answer how a possible act, which really will act more like general legislation regarding resilience in security should be designed and which principles it should employ. We do so on the basis of the knowledge which the economics of security as a field has generated since its inception \cite{Kianpour2021, Wirth2017}, and based on how existing European legal frameworks have succeeded or failed to regulate new areas. 

That being said, security is in no way a new field\footnote{\emph{De vita Caesarum} details encryption, and was written in the year of 120.}, and concrete legislation is certainly welcomed by the academic community as a whole, but the industry and even member states will likely struggle and use all power possible to wiggle out of their potential new responsibilities, maybe with the tools seen elsewhere such as obfuscation\footnote{We have seen this in recent in research such as that by Ferrari \cite{Ferrari2022}, where views and points which are not expressed by the specific group are created, and used to or not to do certain things. This could very well happen here too.}. We will discuss the different ways which the rules could be implemented below, and which one would be the most fitting considering the inherent \emph{goals} which regulating cybersecurity in the best manner possible always will have. After all, such legal mechanisms should not end up as ``security theatre'' \cite{Schneier2003}, and this paper is furthermore inspired by the report 'Standardisation and Certification of Safety, Security and Privacy in the ‘Internet of Things’\footnote{\cite{Maple2017}.} as it contains useful recommendations which we can build further upon.\footnote{And while outside the scope of this article, we agree that digital services like Google Maps should be covered by perhaps a new directive such as a `Digital Service Liability Directive' (or a digital equivalent), to prevent situations where companies like Garmin will be held accountable over its software on its devices through the traditional Product Liability Directive, but Google will not, see \cite[22]{Maple2017}.}


Innovations and development in CS overall, and in security in particular, is heavily created and maintained by researchers. The cycle of invention, then adoption by the industry with disregard for constraints or systems that they perceive to be too expensive for them to bother to fix\footnote{Such as the mountain of glaring security issues with IoT, which as of the time of writing is still not rectified.} \cite{Anderson2005}, mean that the foresight which the original creators and researchers bring into the picture is paramount. With this in mind, this paper focuses on these ideals and goals, and cannot keep ``innovation'' and ``business'' in mind - this we know that the Commission and other researchers are far better at.

We recognise that there already exists EU legislation on cybersecurity through the NIS directive\footnote{Directive 2016/1148, concerning measures for a high common level of security of network and information systems across the Union, [2016] L 194/1.} and the Cybersecurity Regulation\footnote{Regulation 2019/81 on ENISA (the European Union Agency for Cybersecurity) and on information and communications technology cybersecurity certification and repealing Regulation (EU)No 526/2013 (Cybersecurity Act), [2019] L 151/15.}, the latter focused entirely on EU institutions. While the spirit and the idea behind the NIS directive is admirable, it does not put proper security into hard law, but leaves it for guidance and certifications and other soft law measures, without enforcement regimes worthy of how central and important security is\footnote{And with glaring issues from the get go \cite{Schmitz-Berndt2021} in the form of very unclear reporting requirements.}. Furthermore, the NIS directive does not implement the practical tools and procedures needed to make systems resilient, instead merely focusing on strategies, the Cooperation Group, security incident response network, and appointments of national competent authorities\footnote{See the NIS directive, Art 1(2).} - many of which are repeats from other EU legislation. On the other hand, the Cybersecurity Regulation represents a good starting point as to how the best security possible can be established\footnote{The empowerment of ENISA in Chapter II is one good example.}, and great concepts being put into hard law\footnote{Art 46 being an example of a proper development of enforceable certifications, but these may conflict with our ideas in this paper.}.

We build on the literature within cybersecurity and within law that applies to digital technologies with additional concepts from safety engineering, a branch which traditionally was connected to physical equipment or structures which can harm or pose a safety risk to legal or physical entities, finance or health or otherwise. But as the EU itself has already recognised\footnote{See page 10 in the `Guidance on Cybersecurity for medical devices' by the MDCG, \url{https://ec.europa.eu/docsroom/documents/41863}, last accessed 5 May 2022.} , safety is becoming security\footnote{Also noted and expanded upon in \cite[1044 - 1045]{Anderson2020}.}, which means we must have these concepts put on top of everything else at all times\footnote{This is not part of the traditional safety to security discussion, as this involved the traditional meaning of security, malicious actions, versus well intended \cite[182]{Leveson1995}. But cybersecurity does contain this element as well, so it is still worth reading.}

The paper is structured as follows:

Section 2 contains definitions, section 3 comments on possible limitations, while section 4 contains arguments as to which policy option is best for the CRA. Section 5 details how an expanded how the purpose should be understood, and section 6 briefly draws up a possible structure of the CRA. Section 7 contains a list of principles which would suit and empower the CRA, all created on the basis of ideas and grievances which the security community has with current security legislation, and which the CRA would greatly benefit from containing, section 8 touches on future work and section 9 concludes the paper.

We find that the CRA should make use of a horizontal framework as an Act, and should by all means possible actually make security resilience real and not just a lofty goal without hard law. To reach its own goals and live up to its title, this approach is the only one which is appropriate. We furthermore find that the CRA must employ principles which acknowledge that safety is becoming security, meaning that understanding that security is not just about fulfilling certifications and standards, but about really committing and fulfilling its duties in the reality which we all use our smartphones in. Lastly, we suggest a structure for the CRA which will resemble the GDPR\footnote{Regulation (EU) 2016/679 of The European Parliament and of
The Council of 27 April 2016 on the protection of natural persons with regard to the processing of personal data and on the free movement of such data, and repealing Directive 95/46/EC (General Data Protection Regulation) [2016] OJ L 119/1.} and ideas from the proposed AI Act combined with good ideas from EU product legislation broadly, to make for an Act which can go beyond for the sake of everyone's security. 


\section{Definitions}

We start the paper by defining certain terms, both because they are important to understand the rest of the paper, but also to clarify our stance on certain ideas and concepts.

\subsection{Safety or Security?}

We define \emph{safety} as freedom from accident or losses \cite[181]{Leveson1995}, which \emph{security} being freedom from adversarial failures. 


As established in the introduction, safety and security are converging, but we can single out situations where it will only matter in one part or the other, even if these are much much rarer than they used to be\footnote{In our paper on surgical robots, we create a taxonomy which can be used to further understand how much safety and security has merged \cite{Ludvigsen2022}, through the interdisciplinary approach of both law and security. It amounts to a situation where both adversarial and non-adversarial failures will includes safety elements at almost all stages.}. If at any point personal or other information is retrieved through adversarial and non-adversarial means, it will not involve a direct safety element\footnote{The exception to this is through indirect means, like information causing safety issues secondarily.}. Likewise, if any system which is included in the IoT or CPS sphere fails in a physical manner that is not related to its software, and it harms a person or causes financial damage, there is no security element in this either. But everything else will involve a combination of the two. This matters in the context of the CRA, as the scope of the Act will be gigantic, and rightfully so. Even if the IoT had not already become as ubiquitous as it is, the regulation of security on consumer and corporate computers and servers warrants its own regulation.


\subsection{Failure, Error}

Different branches of engineering and analysis of risks or management apply various definitions to these terms, but we see value it clarifying it it further in this context. Anderson uses the system of \emph{fault} causing an \emph{error}, which leads to a \emph{failure} \cite[251]{Anderson2020}, specifically in regards to distributed systems, which will apply to most of the subjects of the CRA. 

On contrast, Leveson uses the idea of all failures being faults, but with some faults not being failures \cite[173]{Leveson1995}, which is very different from above. Furthermore, \emph{unreliability} is the probability of a failure\footnote{Illustrated as the ``Bathtub'' model on page 174 \cite{Leveson1995}.}, and \emph{failure} must then be the event of a system being unable to perform intended function \cite[172]{Leveson1995}. \emph{Failures} are events whereas \emph{errors} are states and therefore static, and caused by design flaws or deviations from the intended state, which leads to failure. Faults which are not failures could include spontaneous or accidental events such as electricity being cut, which would be a fault in a given system, but a failure in the main electric grid. 

A comparison is given in \hyperref[Figure 1]{Figure 1}.

In the context of the CRA, this means that there was a fault, but that it was a fault and a failure outside of these security systems\footnote{Leveson makes a greater argument for dividing failures into three categories and distinguishing between faults \cite[174]{Leveson1995}, which would work great in a different kind of paper, but which could potentially play a role in detailed guidance further on.}. 

\begin{figure}[h]
\caption{Comparison between the two models in regards to failure.}
\centering
\includegraphics[width=\textwidth]{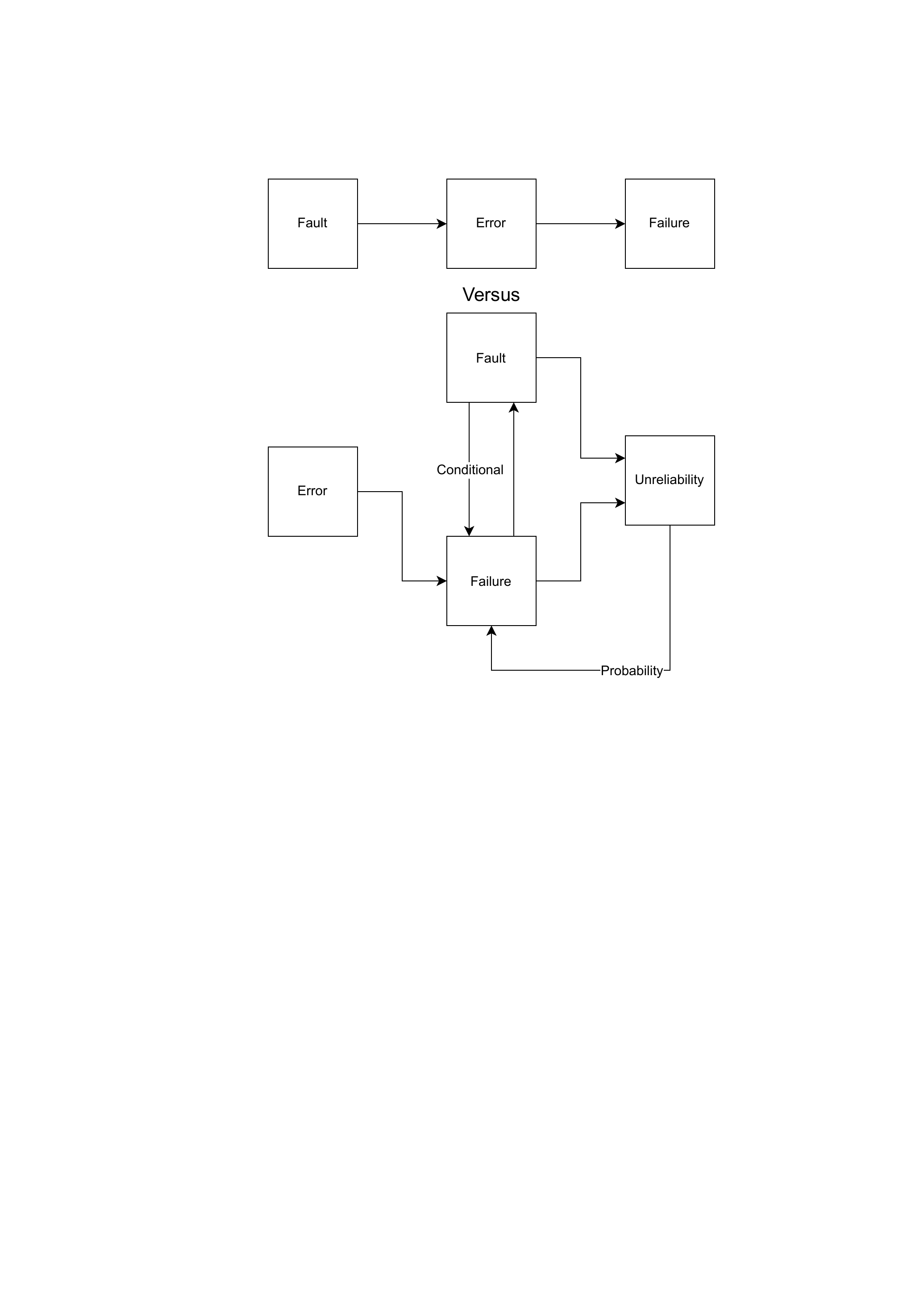}
\end{figure}\label{Figure 1}

For ease of understanding, the first model is the most pragmatic and is the most straight forward, but as safety is becoming security, the complexity and attack venues of complicated physical systems interacting with the digital warrant a combined model at a later date, or perhaps in time with the potential creation of the CRA.


\subsection{Accident and Incident}

\emph{Accidents} in security can be expanded to include any undesired and unplanned event that results in losses. Contrary, \emph{incidents} are near misses that involve no or little loss, but which can turn into accidents later or currently depending on circumstances \cite[175 - 176]{Leveson1995}. Clearly, failures can be both accidents and incidents, but accidents and incidents are not always caused failures of the security system, and there will also be errors (not failures) which can cause accidents and incidents. The distinction matters because the type of event may cause different types of procedure at litigation, errors should be well known due to their static nature, while failures are less expected (but still can be due to their accidental nature), akin to the distinction between systemic and non-systemic issues in product liability \cite[11 - 13]{Ludvigsen2022}. 

\subsection{Hazards}

\emph{Hazard} is defined as a state or conditions of a system combined with the security system and anything around it (faults too), which will lead to an accident \cite[176 - 177]{Leveson1995}. This makes it identifiable and relatively predictable - which is a tradition in safety engineering, to learn on an empirical basis and react from past events and experiences, as was seen with various industrial accidents leading to new legislation in the past. In this sense, a hazard is more like a combination of what was seen in \hyperref[Figure 1]{Figure 1} than a single point which we could add to it.

\subsection{Risk}

\emph{Risk} is the hazard level (severity and likelihood) plus likelihood of hazard leading to accidents and hazard exposure \cite[179]{Leveson1995}. In the context of the CRA, this is the central calculation which must be used to gauge and mitigate or at least lower the damage caused by accidents, but calculating chances for hazards to occur with it is important too. Clearly, the lower the hazard level and likelihood of an accident, the less there should be done, unless the type of accident which this risk could cause is immense. 

The classic example here is the assessment of the Challenger space shuttle, which continue to have consequences to this day \cite{Wright2018}. 

An example from safety (but within a network environment) could be a type of non-adversarial attack which a surgical robot can cause to itself \cite[4]{Ludvigsen2022} to stop functioning for an unspecified amount of time. The hazard level is low. The exposure, time where the hazard affects the system, is low, and the likelihood of the accident happening is likely very low. Still, outside of the risk measurements, the physical consequences of the accident (not calculated here) is gigantic, as the patient could be injured or die from this very rare type of hazard. 

Furthermore, risk is a concept which we assume and or at least make likely, it likelihood after all. This is why we cannot wholly rely on it, but it is clearly the best we have \cite{Nau2001}.

\subsection{Resilience} \label{Resilience}

\emph{Resilience} will then be elements of a system (be it small or massive and based on IoT) which allow \emph{fault detection, fault tolerance, error recovery and failure recovery} \cite[251 - 252]{Anderson2020}. This constitutes the core of the CRA is supposed to do, but as can be seen already here, to understand resilience, you must understand every other part of its being and all types of recovery and detection. None of these are easy, and rely on classic engineering ideas such as redundancy, logs, backups and so on.

Actually implementing resilience is not the same as proposing its existence, akin to calculating risk and actually managing it. The choice of tools depend on which type of defence is needed\footnote{Anderson makes a great overview of some two general issues \cite[252 - 258]{Anderson2020}.} What matters is whether the resilience is anchored by the practical failure tolerance/recovery/preventive systems, have redundancy in the form of backup servers does not matter if the adversary also has hit them with ransomware, or if they keep on initiating attacks on the basis of errors in IoT equipment which the manufacturer is never going to patch. It is therefore paramount to understand resilience as a constant level of readiness, not as an attribute that can be put into a system and then abandoned. 

\section{Potential Limitations}

Within EU-law, there are certain limits as to what member states will have to fulfill and oblige to do. These are limited by the treaties. Furthermore, legislation like NIS or the current Cybersecurity Act will limit how much the future CRA can potentially do, although not by much, due to the relatively soft approaches both types of current legislation have taken\footnote{See \cite{Wallis2020, Ludvigsen2022, Ludvigsen2022a, Wallis2021, Ducuing2021} for examples.}.

Initially though, and since cybersecurity does not yet belong under any of the exclusive competences such as competition from Article 3(1)(b) of the Treaty of the Functioning of the European Union \cite{tfeu}, we must see whether there are limits in regards to the shared competences in Article 4 instead. Article 4(2) is as follows:

\begin{itemize}
\item (a) internal market;
\item (b) social policy, for the aspects defined in this Treaty;
\item (c) economic, social and territorial cohesion;
\item (d) agriculture and fisheries, excluding the conservation of marine biological resources;
\item (e) environment;
\item (f) consumer protection;
\item (g) transport;
\item (h) trans-European networks;
\item (i) energy;
\item (j) area of freedom, security and justice;
\item (k) common safety concerns in public health matters, for the aspects defined in this Treaty
\end{itemize}

We see that security is not contained in (b), (e) and (k). Security plays a role in every product where relevant, as a feature of the market, in agriculture and fisheries (if they use products or systems which need it), in transport on several levels, in the networks, clearly in energy (critical infrastructure angle or not), and is a key feature in freedom/security/justice. Where does this leave the principles below? Safe, exactly because of existing legislation which touches and uses similar enforcement structures, such as the GDPR. 




\section{The 5 Policy Options}


In this section, we comment on each of the suggested approaches which the CRA could make use of.

\subsection{Status Quo}


Standards and certifications regarding cybersecurity are not followed, and contrast the meticulous and insanely accurate implementations of them in industries that make use of technology which can kill\footnote{Such as Electrical Engineering, see \cite[18]{Maple2017}.}. For this reason alone, status quo is not an acceptable solution.

\subsection{Voluntary Measures}

Voluntary measures outside of EU law already exist, and are as the same reasons above, inadequate, but definitely better than status quo.

\subsection{Ad Hoc}

Only a marginal improvement from voluntary measures, ad hoc at least involves the opportunities for changing existing requirements when needed. This approach does not solve the main issue with certifications and standards, but is still preferred over the two above.

\subsection{Mixed Approach}

The mixed approach has two distinctions - firstly, it focuses on a schism between embedded and non-embedded security, but there is no reason to make this distinction, due to how security functions in modern systems. IoT systems at a hospital which are no longer maintained by the vendor, can be the reason why non-embedded systems may be compromised, which makes this schism pointless and potentially a liability, the highest standard possible should be applied to all, and users have the right to make use analogue systems still. The other distinction is how it functions - a horizontal approach with actual security requirements (hard law unlike above), but these are complied with through self-assessments or third party assessment, which would be lower than compliance requirements for many types of products in the EU, such as software in or as medical devices \cite{Ludvigsen2021}. In this sense, even if the risk can vary because of different likelihoods of accidents, because many types of attacks can be applied universally (such as DOS), there is no point in not preventing it at scale, especially if the marginal costs are not that much larger. 

\subsection{Horizontal Framework}

This includes all the ideas above, but where self-assessment is not the standard form, and can either include all software or only ``critical software''. This is a good approach, because it can achieve all the goals which all the options above cannot through enforceability. When building up new legislation to regulate new areas, the EU tends to mimic its biggest successes, which is rather rational and useful. In this situation, taking the best elements out of Regulations like the GDPR seem prudent.

There is crucial difference between the Call for Evidence document and our approach - third party conformity assessment akin to what is seen in product legislation like the Medical Device Regulation\footnote{Regulation (EU) 2017/745 of the European Parliament and of the Council of 5 April 2017 on medical devices, amending Directive 2001/83/EC, Regulation (EC) No 178/2002 and Regulation (EC) No 1223/2009 and repealing Council Directives 90/385/EEC and 93/42/EEC [2017] OJ L117/1. 15.} is not sufficient, leaving the both costly and dangerous task for assessing security should not be left to private parties. Instead, we proposed a dedicated authority in each Member State, as detailed \hyperref[Enforcement]{below}.

\section{The Purpose}


The three purposes stated by the Commission are:

\begin{enumerate}
    
\item To enhance and ensure a consistently high level of cybersecurity of digital products and ancillary services, secured throughout their whole lifecycle proportional to the risks.\footnote{See page 2 - 3 in the Call for Evidence.}
\item Match users to match the security properties of products with their needs, which should protect users from insecure digital products and ancillary services, and incentivise vendors to offer more secure products.
\item To improve the functioning of the internal market by levelling the playing field for vendors of
digital products and ancillary services.\footnote{Purpose 3 is not considered in this paper.} 

\end{enumerate}

This creates at best a 50/50 between market considerations and users. However, users are not alike, thus perhaps specifying and protecting consumers explicitly would be beneficial, but this may also be ideally done in practice by consumer rules in general. 

Different parties have different interests, and this is not expressed well in these purposes - to ensure the first purpose the interests of both private and public users should be considered. Private would appreciate and want a high level of cybersecurity, but would be influenced by Member States or even the Commission when these suggest lowering the level in the form of breaches of encryption\footnote{See the proposed Regulation laying down rules to prevent and combat child sexual abuse \cite{CSAregulationproposal}, which is commented on briefly in a forthcoming article \cite{ludvigsen2022d}.}. This would then violate purpose 1. 

Conversely, public users want high cybersecurity in specific areas such as military or other authorities, while insisting on back door and every means possible to breach purpose 1. 

The CRA must consolidate these differences (within the treaties), which is done with insisting on the high level at all times, which can either be done by acknowledging these weaknesses or distinguishing between these two interests that may not align.

Purpose 2 is supported through several of our proposed principles due to the increase focus on transparency. 






\section{Regulatory Mechanisms and Structure}\label{Enforcement}

In this section, we sketch an enforcement structure.

\subsection{Compliance}

Compliance is dictated by the culture, behaviour and place in society of the subjects which you aim to regulate. Security is implemented everywhere, which means the compliance of rules surrounding it must depend on its context, or so we wish things were. In reality, military intelligence members have smartphones, medical equipment may still use Windows XP, social media platforms use poorly implemented and easily attackable ML models and so forth. Adjusting expectations from assuming we can divide usage into different levels may therefore not be relevant, even if the comparison between critical infrastructure to consumer devices would normally entail different levels of security, it does not have to. Considering the amount of devices that are used by consumers, it should be at least equal, devastating attacks on such devices in bulk may even be akin to attacking critical infrastructure.

Compliance cannot be reached with certificates, as these are not reviewed or otherwise renewed or controlled at a rate which breeds confidence \cite{Anderson2020}, rather the exact opposite. But they create a very strong basis for which hard legal rules to create, and for which reasons. And they can be made in a way where their status is reviewed, making them valuable tools if an active part of the compliance system. 

\subsection{Enforcement}

Due to the central role security has in the lives of most people, either directly in their pockets or through companies or the Member States themselves (security in their systems), parallels to GDPR are quite adequate. This implies strong enforcement, and such enforcement is only possible through similar mechanisms. We therefore clearly recommend national authorities with inspection powers, and requirements for staffing inspired by those from the proposed AI Act\footnote{See \url{https://eur-lex.europa.eu/legal-content/EN/TXT/?uri=CELEX:52021PC0206}, last accessed 24 May 2022.}, and sanctions which are similar to those that can be given in the GDPR, but with the additional layer of right to ban products on the market which is seen in much product legislation. In addition to this, legacy systems which persist, and radiate a certain risk, should be explicitly tackled by these kinds of powers.

\subsection{Guidance}

Good guidance can be enforced via main sources requiring their implementation. This therefore leaves room for specific sectors or specific requirements which this paper cannot predict - akin to the role they play in the MDR. We recommend that this approach is followed, even if security is not only seen through products, but in reality everywhere.

\subsection{Structure}

A short description of how a possible CRA could look would be\footnote{This is done in a straight manner, for the actual CRA the chapters would be divided into further divisions akin to existing EU legislation.}:

\begin{itemize}
    \item Chapter 1, general provisions, necessary definitions. \item Chapter 2, principles of security.
    \item Chapter 3, requirements for security.
    \item Chapter 4, notification and reporting\footnote{Only the necessary rules that NIS 1 and 2 do not loyally cover.}.
    \item Chapter 5, standards and certifications\footnote{Specifically those created and maintained by ENISA.}, risk evaluation and assessments and mitigation systems, ex ante and ex post analysis.
    \item Chapter 6, definitions of national authorities, requirements, obligations, powers and liabilities for them, and rules on forced coordination and cooperation with ENISA and other EU authorities.
    \item Final provisions, detailing interactions with product and other relevant legislation.
    \item Annexes as needed.
\end{itemize}

We illustrate how the CRA could look from the perspective of the manufacturer of a security system in \hyperref[Figure 2]{Figure 2}, which we imply without noting their existence in the figure.


\begin{figure}[h]
\caption{Sketch of how a security system would interact with the proposed idea of the CRA.}
\centering
\includegraphics[width=\textwidth]{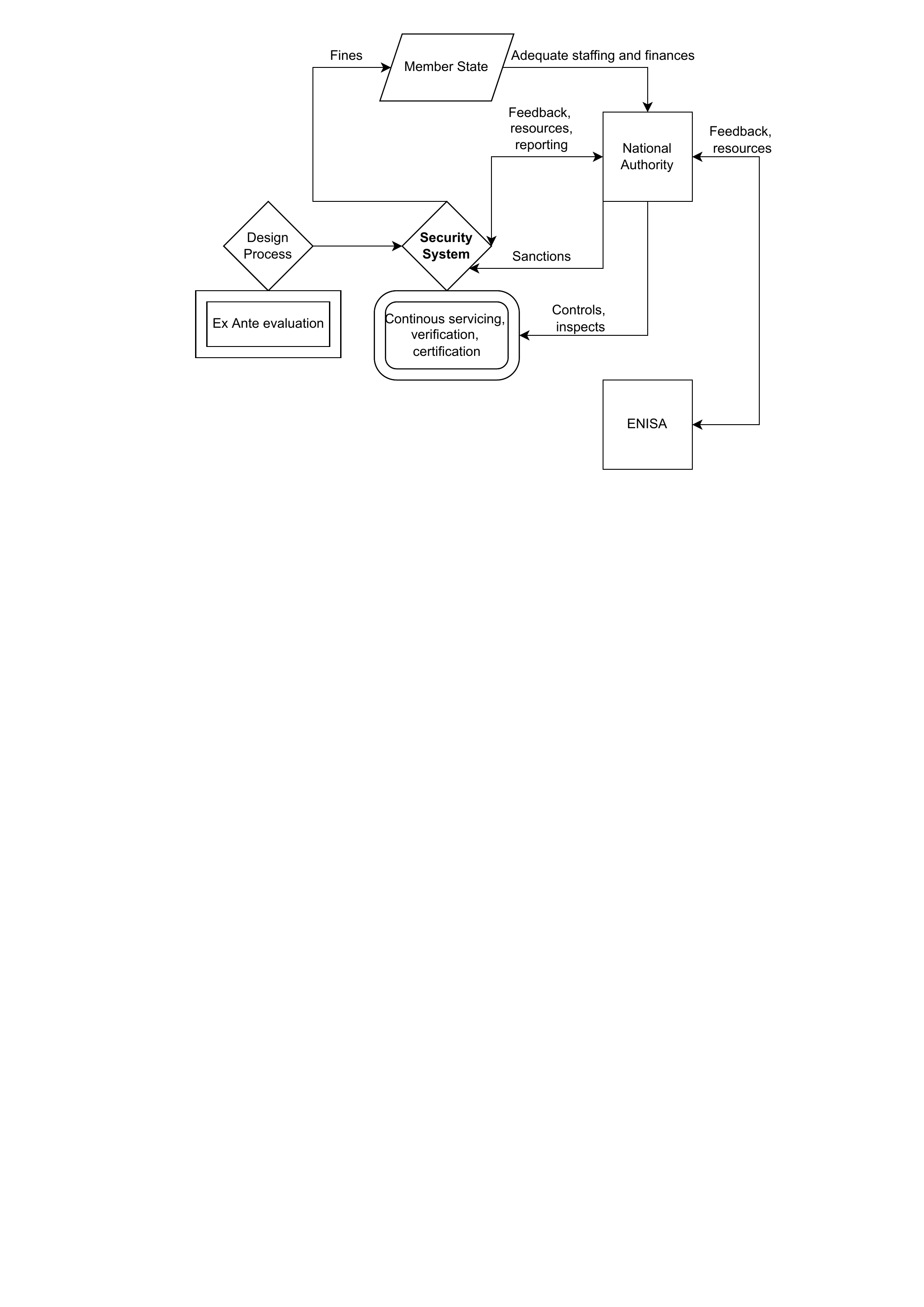}
\end{figure}\label{Figure 2}

\section{Principles}


This paper does not pretend to be able to show all possible principles which could be used to regulate security, but we have selected a big handful which cover contested topics, and which the EU is uniquely placed to regulate due to the Brussels effect \cite{Hadjiyianni2021, Peukert2022}. These take security and the users' seriously, and at the same time offer quite realistic and direct alternatives to the lax structure which exists currently. The order which they are presented fit when they will matter in the life-cycle of a security system, with the earliest in front, and anything mattering later further down, and general principles in between.

\subsection{Ex ante and Ex post assessment}


There is a whole world of literature on how to predict issues with security, both at the design, production and later stages, which is why the CRA must acknowledge and require this.

Security as a field recognises and capable of many things, but because of the economics in it, it follows the same types of production cycles as everyone else. Since many systems and techniques are repeated and recycled because of costs, so will the weaknesses and hazards. This can be ended with CRA, but it must then mandate that these things are assessed, \emph{ex ante}, before the product is put on the market. 

This includes analysis of the open source libraries and other types of freely gotten techniques, which may by themselves contain weaknesses, or which may encounter them later, when the original authors for example change the code, and everyone that uses them automatically updates and perhaps imports the hazards. 

Evaluating code, and what the systems actually do (two very different things), as running code and software on hardware is not the same as security situation as a whole, requires both manual analysis, and perhaps software which can observe obvious weaknesses. These and the way which this kind of ex ante evaluation could performed could be recommended both as a preamble and later as actual requirements with a reference to an Annex that then in details specifies which requirements could be relevant. 

The same can be said for \emph{ex post} evaluations, but these could either be combined with existing market surveillance mechanisms or even in collaboration with a central European agency (like ENISA), so that SMEs would have a much easier time showing security compliance in the long run.

\subsection{Safety and Security by Design}

Inspired by similar ``by design'' principles elsewhere, clearly stating that security should be present at all stages of designing and implementing systems, whether they are purely digital, physical or IoT or CPS, seems prudent and useful. As much as these principles often only amount to statements than only later will be enforced by courts of through public authorities, they still set the standard normatively and can guide and otherwise be on the mind of developers everywhere.

\subsection{Denial of Obscurity}


The understanding of cryptography and cybersecurity in general, should not rest of analogies to trade secrets. There is no research that indicates this being adequate, and no logical argument to be made to support it. Instead, we must focus on what really seems to always be the end result within every branch of cybersecurity - the quality of the system which you construct or the keys or codes you use. 

Kerckhoffs formulated this in 1883, that this quality must be what is chosen over merely keeping it secret \cite{Kerckhoffs1883}. It is called Kerckhoffs' principle. This comes back to a basic assumption of cybersecurity, which is that secrecy is a fallacy. There is always a risk of the measure being guessed or otherwise deduced, therefore focusing on familiar but very robust and perhaps (if possible) unconditionally secure systems is preferred. Including this as either an ultimate requirement or at the very least a goal for cybersecurity in Europe, is therefore highly recommended, and supported by both the great researchers of the past and present \cite{Anderson2020}. 

You are merely rolling the dice and barely buying any time by keeping it secret, because of the exponential complexity required to do so (within any system), compared to using one which inherently is hard to break at all times, regardless of whether 1000s of researchers have attacked and described it\footnote{'The Open Source test' by Raymond confirms that all security is furthermore likely to fit perfectly as free and open source software. See \url{http://www.catb.org/esr/writings/cathedral-bazaar/magic-cauldron/}, last accessed 1 May 2022.}.

There will exist exceptions to certain types of hardware verification mechanisms in for examples CPUs\footnote{Here we specifically refer to enclaves \cite[682 - 685]{Anderson2020}.}, but all of this could be made as an exception and further specified in the Annex - what matters is the concrete attempt to acknowledge a common principle before assessing and seeing the very few exceptions that confirm the rule.


\subsection{Dismissal of Infallibility} 

Technology should not always be perceived as silver bullets. This is very much true for security as well, which makes a preamble or an article that describes why `Titanic syndromes' must be removed or at least something that is considered before new solutions or methods are adopted to the fanfares of marketers and pundits. We have seen this happen both in the past with examples such as escrow keys\footnote{Still heavily in use, including systematically in China \cite{Laskai2021}.} \cite{Blaze1994} and the `unhackable' algorithms in various systems \cite{Anderson1994}, and it is also happening right now, for example via recklessly adopting ML models without realising that this makes the system that uses them more insecure overall\footnote{See examples such as \cite{Fredrikson2015, Papernot2017}.}.

We therefore suggest a statement or a clear rule that forbids notions of infallibility - it is much more coherent and realistic to assume that nothing is infallible and can always suffer adversarial or even non-adversarial failures. This would fit particularly well within the CRA as it acknowledges the whole point or resilience as described earlier.

\subsection{Systems Acknowledgement}


Security is not confined to specific parts of systems (though it can be), more often than not it is part of the entire infrastructure. If so, this means that we too should legislatively acknowledge and be very upfront about this view. We can show why with the following example: A surgical robot makes an unintended movement during operation, and the patient is injured. Further investigation and through standardised digital auditing, it is found that an adversary had made the robot make this one false movement by mimicking the communication between the control system where the surgeon sat and the robot itself, and they did so being physically present and on the network in the room nearby\footnote{Example is constructed from existing research \cite{Bonaci2015a}.}. We then have several layers of the system - the hospital (physical security), the network and the cyberphysical structure of the surgical robot, which is standardised control system to the actuators (the surgical knife). If the CRA only applies to the surgical robot, all the types of adversarial attacks that only make use of the network will not be regulated, which would be against the intention of the entire Act. 

We therefore recommend a statement or part of an article which makes this view concrete - that to cover all types of security, the Act must also cover security in systems as a whole, though perhaps not to the extent where entire councils or countries would be considered one.

\subsection{Full Transparency}

We note that \cite{Maple2017} shows a believable example as to why transparency matters. But not only will transparency between public authorities, CERTs and massive companies be part of the solutions, so will transparency in general. This is why we propose it containing more than one definition:

\emph{One}, transparency in the existing sense, where the sharing and direct cooperation is strengthened to both remove and otherwise document the potential dangers which cannot be shared.

\emph{Two}, transparency regarding which security mechanisms are used and how they work. We remember the principle mentioned above, Denial of Obscurity, and this ties directly into the version of transparency. If we make use of well tested and functioning security mechanisms, everyone should know which, especially consumers, who should be able to understand and see what they make use of.

\subsection{Movement towards a Zero-trust Security Model}

Zero-trust is what EU's own institutions move towards according to the proposal the Cybersecurity Regulation\footnote{Proposal for a regulation of the European Parliament and of the Council
laying down measures for a high common level of cybersecurity at the institutions, bodies, offices and agencies of the Union, 22 March 2022, COM(2022).}, therefore this should be a stated goal in the CRA as well. 

We recommend this being expressed in a preamble, as the costs for enforcing this are higher than traditional external protection barriers (like firewalls, layers in security of a system), or could be a long term goal, expressed in an article, which could be activated at a later date.



\subsection{Cybersecurity Resilience}

The main purpose of the CRA is make this specific term reality, but as defined in \hyperref[Resilience]{above}, resilience requires:

\begin{enumerate}
    \item Fault detection. Assessing risk is not the same as detecting potential faults in systems, this must be separate and separately covered.
    \item Fault tolerance. Even if faults are detected, there may be ways to let them exist in the system without causing errors.
    \item Error recovery. The system has not failed yet, but an error has occurred, and this means it must be designed with this in mind.
    \item Failure recovery. If everything goes wrong, there should still be contingency plans for recovering from the worst possible outcome.
\end{enumerate}

The CRA should have these ideas codified, and then specify in the Annex and guidance how and which tools to use, be it redundancy, code review, internal and external auditing and so on.

\subsection{Enforced Circular Risk Management}

Pressing issues with current models of risk management, such as those used in the MDR, is that as soon as the product is replaced or servicing ends, updates, be it safety or security related, any notion of inspection or compliance ends. This is very different with cybersecurity - many products and systems are continuously used regardless of age. While the risk of this should not be forced onto manufacturers, necessarily, the CRA could keep this in mind in its preamble. 

The second part of this is the ``circular'' part. Security is a never-ending arms race, and this should be expressed in its rules on risk management. Enforcing existing well written guidance such as that by MDCG in the 'Guidance on Cybersecurity for medical devices', thereby making excellent versatile security guidance from soft to hard law, seems incredibly fitting and quicker to implement than creating new terms from scratch. 

\subsection{Dependability}

Like reliability in safety engineering, security in all types of systems may need to be very dependable for longer periods of time. This could also be required of every kind of system, but given that we cannot just changes the economic of security, we have to acknowledge and allow this to a certain degree. But other types of security systems, like those used in critical infrastructure, medical devices, for military purposes and even those we carry around each day, must have degrees of dependability. While this can be expressed within one of the other principles in practice in the CRA, stating that most systems should be build on the basis of reliable and known solutions, and be reliable and usable for when they are needed, would be a great and useful statement to make in the preamble.

\subsection{Hazard Analysis and mitigation or limitation}

A swift way to describe how hazards should be analysed and mitigated or attempted to, could be on the basis of ISO 62366-2. Implementing this standard into hard law would be an easy solution, and it is well known, understood and approved within the community, even if the implementation costs are not small. Controlling circumstances which one is sure will lead to accidents is paramount, and well known within security, for example knowing that using a ML model will lead to certain vulnerabilities related to classifiers \cite{Tramer2016}, or knowing that extra air gaps or specialised coating is needed for a special server, as an adversary could guess what is happening inside with various passive attacks (including electro-magnetic leakage) \cite{Sayakkara2021}.

\subsection{Liability}

While not a single principle, delegating liability for cybersecurity should not be left analysed or otherwise relegated to default notions. This is due to the complexity of who creates and inspires or otherwise writes the code or physical features in use. To understand this, we need to realise that cybersecurity like any other part of CS relies on open source or otherwise freely available code and ideas, both in a copyright sense or in an expanded manner. The ideas and concepts of academics, freelancers, amateurs or the highest paid internal researchers in the world may have contributed to whatever security that then fails you, and who would you want to keep liable? We know from European Product Liability, that it would under most circumstances be relegated to the manufacturer of the product, but what if they only licensed the part that failed in your lawsuit? We can solve these issues through insurance, and in that case the manufacturer does not care, they will take that fight with the insurance company or sue the third party that produced their faulty code. We believe this approach should continue, but there is good reason to illustrate and otherwise heavily acknowledge, in the text or in the preambles, that liability may split depending on the circumstances, and that this should be reflected in how the future Act is understood. 

Mechanisms which encourage manufacturers or others that may kept liable to publicly announce which freely available ideas or code should be avoided, should be encouraged by the member states and perhaps systematised.

\subsection{A Clear Reporting Regime}

While reports and communication has so far been left up to vague rules and concepts. To quickly solve this, IEC 29147/2014 and IEC 30111 can both be cemented into law. Combined with the extensive pool of experience which the industry has from having both handled bug bounties on top of receiving information from CERTS or authorities at large, and we can create a combination of them all that can encompass everything at once.\footnote{As suggested by \cite[50 - 52]{Maple2017}.} 

\subsection{Enforcement of Certifications and Standards} 

We should not throw out the baby with the bathwater. The Act gives an opportunity to not only create a basis for both safety and security testing, but also enforcing native existing standards and certifications. Both of these only matter if they are constantly adhered to, and sanctioning if non-compliant is therefore necessary. The EU can chose two paths within this: Enforce and create the rules nationally or a hybrid model with a central authority like ENISA that standardises some measures and leaves the rest up to the member states. Either will likely function, combined with the ideas put for the in the Cybersecurity Act.

\subsection{Mandated Verification of Security}

Having security, acquiring certificates and otherwise showing what you think you posses in security terms is neat, but what is even more important is being able to verify the existence of the security, and show its performance in practice. Penetration testing is currently the norm, even if it is not mandated by hard law so far. This could be made possible with the CRA, and would not require massive sweeping rules to become reality as the industry already possess the necessary tools to perform the tests regularly. This could be part of the life-cycle of a system, and be either specified as an article or defined narrowly in an Annex.

\subsection{Continuous Servicing}

As Anderson notes \cite{Anderson2020}, the Sales of Goods Directive\footnote{Directive 2019/771 of the European Parliament and of the Council of 20 May 2019 on certain aspects concerning contracts for the sale of goods, amending Regulation (EU) 2017/2394 and Directive 2009/22/EC, and repealing Directive 1999/44/EC, OJ L 136/28.} mandates that security is maintained 2 years after purchase, or for longer if expected by the customer. Firstly, this is a directive, which means it may lead to fragmentation and otherwise may be used in creative or non-conform ways within the different member states. To prevent circumvention of this, the CRA could mandate that the latter takes precedent over the former, as security is also about protecting individuals and adhering to their expectations\footnote{Human and Computer Interaction is its own field for a reason, and much great research has been done on the interplay between security and humans in recent years.}. But we do agree that the CRA should not supercede this Directive or perhaps its child in a Regulation or Act form, which is why the notion of ``continous service'' could be well expressed in a preamble to really settle the debate, at least regarding security going forward (both for consumers and others).


This includes penetration testing and similar functions. These have been notoriously hard to enforce, but this regulation gives an opportunity for the EU to rectify this.\footnote{See \url{https://www.enisa.europa.eu/publications/good-practices-for-an-eu-ics-testing-coordination-capability}, last accessed 7 May 2022 for more, note the date of publication.}



We suggest principles which include empowering and increasing security for European Citizens, transparency and dependability, concepts and rules for the entire life of the systems, verification and even legal issues. The commonality between all of them is their focus on pragmatism - in what is easy to implement legislatively and in practice, as we have come very far from the systems and resources of the past.

\section{Future Work}

Regulating a pervasive yet extremely important area like cybersecurity requires many types of additional research, and if any event is enough to cause such a boom, the bountiful amount of law that regulates it in the EU should be. We need all types, both in terms of potential impact on businesses, traditional security, how it will be dealt with through litigation and through lenses such as justice. Why should some types of security (critical infrastructure) be better than in devices of those with most or least, when this kind of good security is rather easy to implement everywhere? The answer probably lies somewhere with the pool of great literature surrounding the interests of both Member States and the EU as such regarding digital surveillance.

\section{Concluding remarks}

The idea of a CRA may seem more blue skies in this paper than it may end up becoming. Regulating security in general often feels like a Sisyphean task, but as Moore's law slows, there is perhaps potential to finally get the appropriate energy and overview of what is necessary to do. 

In this paper, we suggest a range of principles which can improve security, its reputation and even potentially rights for citizens and users of it. This means a net gain for all parties. We also sketch how the CRA could look and why it should be horizontal framework above anything else. Emphasising strong regulators, good and regularly audited certifications, and codifying the strong principles would entail a new era for security both in the Europe and the world, and the first big step towards properly (but not oppressively) regulating security in the long run.

\printbibliography

@article{Blaze1994,
author = {Blaze, Matt},
doi = {10.1145/191177.191193},
isbn = {0897917324},
issn = {15437221},
journal = {Proceedings of the ACM Conference on Computer and Communications Security},
pages = {59--67},
title = {{Protocol failure in the Escrowed Encryption Standard}},
year = {1994}
}

@article{Hadjiyianni2021,
author = {Hadjiyianni, Ioanna},
doi = {10.1093/ojls/gqaa042},
issn = {14643820},
journal = {Oxford Journal of Legal Studies},
number = {1},
pages = {243--264},
title = {{The European Union as a Global Regulatory Power}},
volume = {41},
year = {2021}
}

@article{Sayakkara2021,
author = {Sayakkara, Asanka P. and Le-Khac, Nhien An},
doi = {10.1109/ACCESS.2021.3051921},
issn = {21693536},
journal = {IEEE Access},
pages = {13237--13247},
title = {{Forensic insights from smartphones through electromagnetic side-channel analysis}},
volume = {9},
year = {2021}
}

@article{Wright2018,
author = {Wright, Jan Folkmann},
doi = {10.1080/13669877.2016.1235605},
issn = {14664461},
journal = {Journal of Risk Research},
number = {6},
pages = {710--724},
publisher = {Routledge},
title = {{Risk management; a behavioural perspective}},
url = {http://dx.doi.org/10.1080/13669877.2016.1235605},
volume = {21},
year = {2018}
}

@article{Anderson2005,
author = {Anderson, Ross},
journal = {Perspectives on free and open source software},
pages = {127--142},
title = {{Open and Closed Systems are Equivalent (that is, in an ideal world)}},
url = {http://www.cl.cam.ac.uk/{~}rja14/Papers/toulousebook.pdf},
year = {2005}
}

@article{Anderson1994,
author = {Anderson, Ross J.},
doi = {10.1007/3-540-58618-0_67},
journal = {Esorics},
pages = {231--245},
title = {{Liability and computer security: Nine principles}},
year = {1994}
}

@techreport{Laskai2021,
author = {Laskai, Lorand and Segal, Adam},
booktitle = {International Encryption Brief},
institution = {Carnegie Endowment for International Peace},
number = {March},
title = {{The Encryption Debate in China: 2021 Update}},
url = {https://carnegieendowment.org/files/202104-Germany{\_}Country{\_}Brief.pdf},
year = {2021}
}

@article{Ferrari2022,
author = {Ferrari, Valeria},
doi = {10.1016/j.clsr.2022.105687},
issn = {0267-3649},
journal = {Computer Law {\&} Security Review},
publisher = {Elsevier Ltd},
title = {{The platformisation of digital payments : The fabrication of consumer interest in the EU FinTech}},
url = {https://doi.org/10.1016/j.clsr.2022.105687},
volume = {45},
year = {2022}
}

@article{Kianpour2021,
author = {Kianpour, Mazaher and Kowalski, Stewart James and {\O}verby, Harald},
issn = {16130073},
journal = {CEUR Workshop Proceedings},
pages = {46--65},
title = {{Multi-paradigmatic approaches in cybersecurity economics}},
volume = {3016},
year = {2021}
}

@article{Wirth2017,
author = {Wirth, Axel},
doi = {10.2345/0899-8205-51.s6.52},
issn = {19435967},
journal = {Biomedical Instrumentation and Technology},
number = {Horizons},
pages = {52--59},
pmid = {29161105},
title = {{The economics of cybersecurity}},
volume = {51},
year = {2017}
}

@unpublished{ludvigsen2022d,
author = {Ludvigsen, Kaspar Rosager and Nagaraja, Shishir and Daly, Angela},
title = {{YASM: Yet Another Surveillance Mechanism}},
year = {2022}
}

@misc{CSAregulationproposal,
author = {Commission, European},
title = {{Proposal for a Regulation of the European Parliament and of the Council laying down rules to prevent and combat child sexual abuse}},
url = {https://ec.europa.eu/home-affairs/system/files/2022-05/Proposal for a Regulation laying down rules to prevent and combat child sexual abuse{\_}en{\_}0.pdf},
year = {2022}
}

@article{Schmitz-Berndt2021,
author = {Schmitz-Berndt, Sandra and Schiffner, Stefan},
doi = {10.1080/13600869.2021.1885103},
issn = {13646885},
journal = {International Review of Law, Computers and Technology},
number = {2},
pages = {101--115},
publisher = {Taylor {\&} Francis},
title = {{Don't tell them now (or at all)–responsible disclosure of security incidents under NIS Directive and GDPR}},
url = {https://doi.org/10.1080/13600869.2021.1885103},
volume = {35},
year = {2021}
}

@article{Ducuing2021,
author = {Ducuing, Charlotte},
doi = {10.1016/j.clsr.2020.105514},
issn = {02673649},
journal = {Computer Law and Security Review},
number = {November 2019},
pages = {105514},
publisher = {Elsevier Ltd},
title = {{Understanding the rule of prevalence in the NIS directive: C-ITS as a case study}},
url = {https://doi.org/10.1016/j.clsr.2020.105514},
volume = {40},
year = {2021}
}

@article{Wallis2021,
author = {Wallis, Tania and Johnson, Chris and Khamis, Mohamed},
issn = {13142119},
journal = {Information {\&} Security: An International Journal},
number = {1},
pages = {36--68},
title = {{Interorganizational Cooperation in Supply Chain Cybersecurity: A Cross-Industry Study of the Effectiveness of the UK Implementation of the NIS Directive}},
url = {https://isij.eu/article/interorganizational-cooperation-supply-chain-cybersecurity-cross-industry-study},
volume = {48},
year = {2021}
}

@misc{tfeu,
author = {Union, European},
pages = {47--390},
publisher = {Official Journal of the European Union},
title = {{The Treaty on the Functioning of the European Union}},
url = {https://eur-lex.europa.eu/resource.html?uri=cellar:2bf140bf-a3f8-4ab2-b506-fd71826e6da6.0023.02/DOC{\_}2{\&}format=PDF},
year = {2012}
}

@article{Ludvigsen2022a,
author = {Biasin, Elisabetta and Kamenja{\v{s}}evi{\'{c}}, Erik},
journal = {International Cybersecurity Law Review},
title = {{The AI Act proposal, the reform of the NIS Directive and the new medical devices' cybersecurity challenges in the European Union}},
year = {2022}
}

@techreport{Maple2017,
author = {Leverett, Eireann and Clayton, Richar and Anderson, Ross},
booktitle = {European Commission},
doi = {10.2760/47559},
isbn = {9789279778636},
title = {{Standardisation and Certification of Safety, Security and Privacy in the ‘Internet of Things'}},
year = {2017}
}

@article{Kerckhoffs1883,
author = {Kerckhoffs, Auguste},
journal = {Journal des sciences militaires},
pages = {5--83},
title = {{La cryptographie militaire}},
url = {http://www.petitcolas.net/fabien/kerckhoffs/},
volume = {IX},
year = {1883}
}

@article{Nau2001,
author = {Nau, Robert F},
journal = {Theory and Decision},
pages = {89--124},
title = {{DE FINETTI WAS RIGHT: PROBABILITY DOES NOT EXIST}},
url = {https://link.springer.com/article/10.1023/A:1015525808214{\#}citeas},
volume = {51},
year = {2001}
}

@book{Anderson2020,
author = {Anderson, Ross},
publisher = {John Wiley {\&} Sons},
title = {{Security engineering: a guide to building dependable distributed systems}},
year = {2020}
}

@book{Schneier2003,
author = {Schneier, Bruce},
publisher = {Copernicus books},
title = {{Beyond fear: Thinking sensibly about security in an uncertain world}},
year = {2003}
}

@article{Papernot2017,
archivePrefix = {arXiv},
arxivId = {1602.02697},
author = {Papernot, Nicolas and McDaniel, Patrick and Goodfellow, Ian and Jha, Somesh and Celik, Z. Berkay and Swami, Ananthram},
doi = {10.1145/3052973.3053009},
eprint = {1602.02697},
isbn = {9781450349444},
journal = {ASIA CCS 2017 - Proceedings of the 2017 ACM Asia Conference on Computer and Communications Security},
pages = {506--519},
title = {{Practical black-box attacks against machine learning}},
year = {2017}
}

@article{Peukert2022,
author = {Peukert, Christian and Bechtold, Stefan and Batikas, Michail and Kretschmer, Tobias},
doi = {10.1287/mksc.2021.1339},
isbn = {0000000308335},
issn = {0732-2399},
journal = {Marketing Science},
number = {March},
title = {{Regulatory Spillovers and Data Governance: Evidence from the GDPR}},
year = {2022}
}

@article{Ludvigsen2022,
author = {Ludvigsen, Kaspar Rosager and Nagaraja, Shishir},
doi = {10.1016/j.clsr.2022.105656},
issn = {02673649},
journal = {Computer Law and Security Review},
publisher = {Elsevier Ltd},
title = {{Dissecting liabilities in adversarial surgical robot failures: A national (Danish) and EU law perspective}},
url = {https://doi.org/10.1016/j.clsr.2022.105656},
volume = {44},
year = {2022}
}

@article{Ludvigsen2021,
author = {Ludvigsen, Kaspar and Nagaraja, Shishir and Daly, Angela},
doi = {10.1017/err.2021.45},
issn = {1867-299X},
journal = {European Journal of Risk Regulation},
pages = {1--16},
title = {{When Is Software a Medical Device? Understanding and Determining the “Intention” and Requirements for Software as a Medical Device in European Union Law}},
year = {2021}
}

@article{Wallis2020,
author = {Wallis, Tania and Johnson, Chris},
doi = {http://dx.doi.org/10.1109/CyberSA49311.2020.9139641},
journal = {International Conference on Cyber Situational Awareness, Data Analytics and Assessment (CyberSA)},
pages = {49--58},
title = {{Implementing the NIS Directive, driving cybersecurity improvements for Essential Services}},
year = {2020}
}

@book{Leveson1995,
author = {Leveson, Nancy G.},
edition = {1.},
isbn = {0-201-11972-2},
pages = {680},
publisher = {Addison-Wesley Publishing Company, Inc.},
title = {{Safeware: System Safety and Computers}},
year = {1995}
}

@article{Bonaci2015a,
archivePrefix = {arXiv},
arxivId = {1504.04339},
author = {Bonaci, Tamara and Herron, Jeffrey and Yusuf, Tariq and Yan, Junjie and Kohno, Tadayoshi and Chizeck, Howard Jay},
eprint = {1504.04339},
pages = {1--11},
title = {{To Make a Robot Secure: An Experimental Analysis of Cyber Security Threats Against Teleoperated Surgical Robots}},
url = {http://arxiv.org/abs/1504.04339},
year = {2015}
}

@article{Fredrikson2015,
author = {Fredrikson, Matt and Jha, Somesh and Ristenpart, Thomas},
doi = {10.1145/2810103.2813677},
isbn = {9781450338325},
issn = {15437221},
journal = {Proceedings of the ACM Conference on Computer and Communications Security},
pages = {1322--1333},
title = {{Model inversion attacks that exploit confidence information and basic countermeasures}},
volume = {2015-Octob},
year = {2015}
}

@unpublished{Tramer2016,
archivePrefix = {arXiv},
arxivId = {1609.02943},
author = {Tram{\`{e}}r, Florian and Zhang, Fan and Juels, Ari and Reiter, Michael K. and Ristenpart, Thomas},
eprint = {1609.02943},
isbn = {9781931971324},
title = {{Stealing Machine Learning Models via Prediction APIs}},
url = {http://arxiv.org/abs/1609.02943},
year = {2016}
}

\end{document}